\documentclass[aps,prb,showpacs,amsmath,amssymb,twocolumn]{revtex4-1}
\usepackage{amsmath}
\usepackage{bm}    
\usepackage{graphicx}         
\usepackage{multirow}
\usepackage{float}

\begin{document}

\newcommand{\bn}{{\bf n}}
\newcommand{\bp}{{\bf p}}   
\newcommand{\br}{{\bf r}}
\newcommand{\bk}{{\bf k}}
\newcommand{\bv}{{\bf v}}
\newcommand{\brho}{{\bm{\rho}}}
\newcommand{\bj}{{\bf j}}
\newcommand{\wk}{\omega_{\bf k}}
\newcommand{\nk}{n_{\bf k}}
\newcommand{\eps}{\varepsilon}
\newcommand{\la}{\langle}
\newcommand{\ra}{\rangle}
\newcommand{\be}{\begin{equation}}
\newcommand{\ee}{\end{equation}}
\newcommand{\intl}{\int\limits_{-\infty}^{\infty}}
\newcommand{\dE}{\delta{\cal E}^{ext}}
\newcommand{\SE}{S_{\cal E}^{ext}}
\newcommand{\dsp}{\displaystyle}
\newcommand{\phit}{\varphi_{\tau}}
\newcommand{\p}{\varphi}
\newcommand{\cL}{{\cal L}}
\newcommand{\dphi}{\delta\varphi}
\newcommand{\dbj}{\delta{\bf j}}
\newcommand{\lra}{\leftrightarrow}
\newcommand{\comment}[1]{}
\renewcommand{\eqref}{\ref}

\title{AC Response of the Edge States in a Two-Dimensional Topological Insulator  Coupled to a Conducting Puddle }

\author{K. E. Nagaev}

%\mail{e-mail \textsf{ nag@cplire.ru}}

\affiliation{Kotelnikov Institute of Radioengineering and Electronics, Mokhovaya 11-7, Moscow, 125009 Russia}

%\keywords{Topological insulator, ac conductance, spin-flip scattering }
%\PACS  73.23.2b \sep 72.70.1m \sep 73.50.Td

\begin{abstract}
We calculate an AC response of the edge states of a two-dimensional topological insulator, which
can exchange  electrons with a conducting puddle in the bulk of the insulator. This exchange leads to finite
corrections to the response of isolated edge states both at low and high frequencies. By comparing these corrections,
one may determine the parameters of the puddle.
\end{abstract}

\maketitle

\section{Introduction}
A signature of two-dimensional (2D) topological insulators is the existence of helical edge electronic states that
propagate in clockwise and counterclockwise directions. As the projection of electron spin is locked to the direction
of its motion, the electron can only change both of them simultaneously and cannot be backscattered by non-magnetic impurities or 
phonons as in conventional conductors. Therefore it was theoretically 
predicted that in the absence of spin-flip scattering, a pair of helical edge states should have a universal value
of conductance $e^2/h$, no matter how long they are \cite{Hasan10}. However experiments revealed that actual values of 
conductance were much smaller. In papers \cite{Konig07,Roth09} reporting measurements on HgTe/CdTe quantum wells, the 
conductance of 1 $\mu$m-long edge states 
was 10\% smaller than expected. Some other paper reported a decrease of conductance by two orders of magnitude 
\cite{Gusev11,Grabecki13}. A similar suppression of conductance was found in InAs/GaSb/AlSb heterostructures \cite{Du15,Knez14}. 
In all experiments, it had a very weak temperature dependence.

So far, there was no satisfactory explanation of these facts despite a large number of theoretical papers in this field. First
the conductance suppression was attributed to spin-flip scattering of electrons by magnetic impurities \cite{Maciejko09},
but it appeared shortly that axially symmetric impurities do not contribute to the dc resistance because of conservation of total
spin of the electrons and impurities \cite{Tanaka11}. To avoid this conservation, the authors of Ref. \cite{Altshuler13} assumed 
that the magnetic impurities have a random anisotropy in the plane of the insulator. They obtained that such impurities would lead 
to the Anderson localization of the edge states and an exponential decrease of the conductance with the length of the sample,
while its experimental values are inversely proportional to its length like in diffusive conductors. This could take place if along
with anisotropic impurities there were sufficiently strong dephasing processes. However the estimates show that the dephasing length 
is much larger than the distance between the probes \cite{Gusev11,Grabecki13,Tikhonov15}, which makes this mechanism unlikely.

Another possibility is that conducting puddles are for\-med in the bulk of the insulator because of potential fluctuations due to
randomness of impurity doping and electrons from the edge states are captured into these puddles,  as suggested by Vayrynen et al.
\cite{Vayrynen13}. The authors found that together with Coulomb interaction, this resulted in a suppression of the conductance, but
its strong temperature dependence did not agree with experiments. Nevertheless scanning-gate experiments \cite{Konig13} suggest that
the suppression arises from well-localized discrete objects near the edges.

Recently, it was suggested that the suppression of conductance may arise from the tunnel coupling between the edge states
and conducting puddles of relatively large size that have a continuous energy spectrum and allow a two-dimensional motion
of electrons in them \cite{Essert15,Aseev16}. The impurity scattering in the puddles combined with spin-orbit coupling may 
result in a temperature-independent spin relaxation  of electrons via the Elliott--Yafet \cite{Elliott54} or Overhauser 
mechanism \cite{Overhauser53}, see Ref. \cite{Zutic04} for a review. The existence of these puddles will lead to an effective
backscatttering of electrons. In particular, it was shown in \cite{Aseev16} that
even one puddle could reduce the conductance by half if the tunnel coupling and spin-flip scattering in the puddle are 
sufficiently strong. However the conductance depends on both of these quantities and therefore it is difficult to extract them
from measurements of dc current. In this paper, we present calculations of a frequency-dependent response of  a pair of edge 
states coupled to a conducting puddle. By comparing the low- and high-frequency conductances, one can determine the parameters
of the puddle and judge upon the applicability of this model.

\section{Model and general equations}

\begin{figure}
  \centering
  \includegraphics[width=0.9\columnwidth]{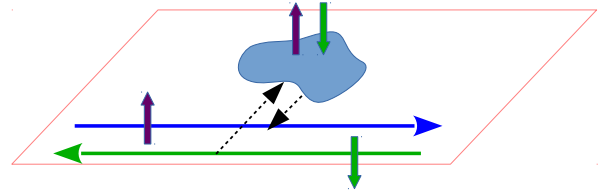}
  \caption{(color online). A pair of edge states tunnel-coupled to a conducting puddle in the bulk
  of the 2D topological insulator. The electron spin in the edge states is locked to the direction of motion, but
  the electrons in the puddle can flip it without restrictions.}
  \label{fig1}
\end{figure}

Consider a pair of helical edge states with linear dispersion $\eps_p=|p|\,v$ that connect the electron reservoirs,
which are kept at externally controllable voltages.  Each of the two directions of the electron momentum is locked to
a definite spin projection, which is labeled by $\sigma=\pm 1$. 
% For simplicity, the interaction between the electrons in these states is neglected. 
The edge states are tunnel-coupled with electron or hole puddles that 
are formed in the bulk of the insulator because of large-scale potential fluctuations. We also assume that these 
puddles are sufficiently large to have a continuous spectrum and that the  electrons in the puddles are also subject to a spin relaxation because of spin-orbit processes.  

For simplicity, the interaction between the electrons in the edge states is neglected, as well as their interaction with the electrons in the puddle.

Hence the 
distribution functions $f_{\sigma}(x,\eps,t)$ in the edge states obey the equation  \cite{Lunde12} 
\begin{multline}
 \left(
  \frac{\partial}{\partial t} + {\sigma}v\,\frac{\partial}{\partial x}
 \right) f_{\sigma}(x,\eps,t)
\\
 =
 -\Gamma(x) \left[ f_{\sigma}(x,\eps,t) - F_{\sigma}(\eps,t) \right]
 -e\,\frac{\partial u}{\partial t}\,\frac{\partial f_{\sigma}}{\partial\eps},
 \label{f-eq1}
\end{multline}
where $\Gamma(x)$ is the rate of electron tunneling from point $x$ to the puddle, $F_{\sigma}(\eps, t)$  is 
the spin-dependent distribution function of electrons in the puddle, and $u(x,t)$ is the electric potential.  As the 
conductance of the puddle is much higher 
than that of the edge states, this distribution functions is  spatially uniform inside it and obeys the equation
\begin{multline}
 \frac{\partial F_{\sigma}}{\partial t} 
 + \frac{1}{h\nu_p}\int dx\,\Gamma(x)\,[F_{\sigma}(\eps,t) - f_{\sigma}(x,\eps,t)] 
\\
 + \frac{1}{2\tau_s}\,(F_{\sigma} - F_{-{\sigma}})
 = -e\,\frac{dU}{dt}\,\frac{\partial F_{\sigma}}{\partial\eps},
 \label{F-eq1}
\end{multline}
where $\nu_p$ is the number of states in the puddle per unit energy, $\tau_s$ is 
the spin-relaxation time, and $U$ is the electrical potential of the puddle.
In its turn, the time derivatives of $u$ and $U$ may be obtained through electric capacity of the edge state per unit length
$c$, the puddle capacity $C$ and the charge-balance equations
\begin{gather}
 \frac{\partial u}{\partial t} =-\frac{e}{c} \sum_{\sigma} \int\frac{d\eps}{hv}\,
  \left[ \sigma v\,\frac{\partial f_{\sigma}}{\partial x} + \Gamma\,(f_{\sigma} - F_{\sigma}) \right],
\\
 \frac{dU}{dt} = \frac{1}{C}\,\frac{dQ}{dt}
 = \frac{e}{C}\sum_{\sigma} \int d\eps \int_0^L dx\,\frac{\Gamma(x)}{hv}\,
 (f_{\sigma} - F_{\sigma}).
 \label{U-eq1}
\end{gather}

As $\nu_p$ and $\Gamma$ may be considered as energy-independent near the Fermi level, it is convenient to 
introduce the integrated quantities
\begin{gather}
 n_{\sigma}(x,t) = \int \frac{d\eps}{hv}\,[f_{\sigma}(x,\eps,t) - f_0(\eps)],
 \label{n-eq1}\\
 N_{\sigma}(t) = \int d\eps\,\nu_p\,[F_{\sigma}(\eps,t) - f_0(\eps)],
 \label{N-eq1}
\end{gather}
where $f_0(\eps)$ is the equilibrium Fermi distribution. Note that these are not the total electron concentrations
because they take into account only the changes of electron number near the Fermi level and do not include the 
shifts of the bottom of the conduction band that result from the oscillating electric potential. 
One may exclude the 
quantity $u$  from Eq. \eqref{f-eq1} to obtain the equation for $n_{\sigma}$ in the form
\begin{multline}
 \frac{hv\,\partial n_{\sigma}/\partial t - (e^2/c)\,\partial n_{-\sigma}/\partial t}{hv + e^2/c}
\\
 +\sigma v\,\frac{\partial n_{\sigma}}{\partial x} 
 + \Gamma\,n_{\sigma}
 = \frac{\Gamma}{hv\nu_p}\,N_{\sigma}.
 \label{n-eq2}
\end{multline}
Furthermore, it is convenient to separate $N_{\sigma}$ into the char\-ge and spin parts $N_Q = N_{+} + N_{-}$ and 
$N_S = N_{+} - N_{-}$. By adding and subtracting Eqs. \eqref{N-eq1} for $\delta N_{+}$ and $\delta N_{-}$ and making
use of Eq. \eqref{U-eq1}, one obtains the equations for these quantities in the form
\begin{multline}
 \left[ 
    \frac{\partial}{\partial t} 
    + \left( 1 + 2\,\frac{e^2\nu_p}{C} \right) \frac{\p_L}{h\nu_p}
 \right] N_Q
\\
 = \left( 1 + 2\,\frac{e^2\nu_p}{C} \right) \int_0^L dx\,\Gamma(x)\,(n_{+} + n_{-})
 \label{N_Q-eq1}
\end{multline}
and
\be
 \left( \frac{\partial}{\partial t} + \frac{1}{\tau_s} + \frac{\p_L}{h\nu_p} \right) N_S
 = \int_0^L dx\,\Gamma(x)\,(n_{+} - n_{-}),
 \label{N_S-eq1}
\ee
where the notation
\be
 \p_L = \int_0^L dx\,\Gamma(x)/v
 \label{phi_L}
\ee
denotes the dimensionless tunnel-coupling strength. This system of equations must be solved together with the bo\-undary conditions 
\be
 n_{+}(0) = \frac{eu(0)}{hv},
 \quad
 n_{-}(L) = \frac{eu(L)}{hv},
\label{n-boundary} 
\ee 
and the current at point $x$ can be calculated as
\be
 I(x,t) = ev\,[n_{+}(x,t) - n_{-}(x,t)].
 \label{I-eq1}
\ee

\section{AC response}
Calculate now the linear response of the system. We assume that the voltage drop with frequency $\omega$ is 
symmetrically applied to the terminals, i. e. 
$u(0)=\frac{1}{2}\,V\exp(-i\omega t)$ and $u(L)=-\frac{1}{2}\,V\exp(-i\omega t)$. 
Estimates show
that for the edge states in HgTe quantum wells, $hv$ and $e^2/c$ are of the same order of magnitude. Therefore 
the terms with time derivatives in Eqs. \eqref{n-eq2} are much smaller than the ones with spatial derivatives and may be omitted
if we restrict ourselves to $\omega \ll v/L$. Using the boundary conditions Eq. \eqref{n-boundary}, one may write
the solutions of these equations in the form
\begin{align}
 n_{+}(x) = \frac{eV}{2hv}\,K(x,0) &
\nonumber\\
 + \frac{N_Q + N_S}{2hv\nu_p} & \int_0^x \frac{dx'}{v}\,\Gamma(x')\,K(x,x'),
 \label{n+-eq2}\\
 n_{-}(x) = -\frac{eV}{2hv}\,K(L,x) &
\nonumber\\
 + \frac{N_Q - N_S}{2hv\nu_p} & \int_x^L \frac{dx'}{v}\,\Gamma(x')\,K(x',x),
 \label{n--eq2}
\end{align}
where
\be
 K(x,x') =\exp\!\left\{-\int_{x'}^x \frac{dx''}{v}\,\Gamma(x'')\right\}.
 \label{K-eq1}
\ee
\begin{figure}
  \centering
  \includegraphics[width=0.9\columnwidth]{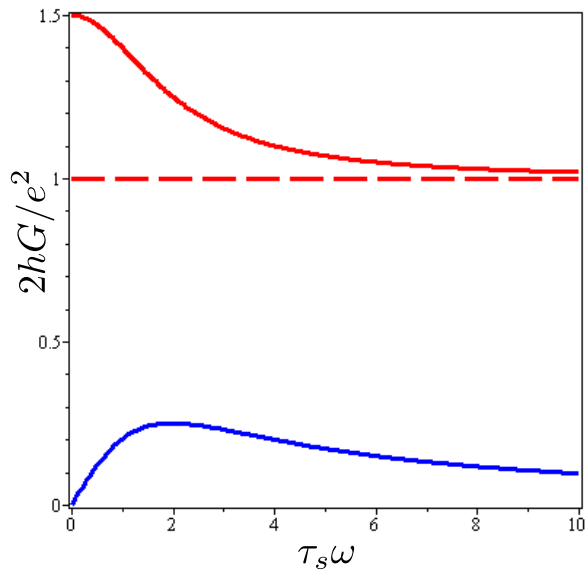}
  \caption{(color online). The real part (red curve) and imaginary part (blue curve) of the conductance vs. frequency for 
  $\p_L=20$ and  $h\nu_p/\tau_s=1$.}
  \label{fig2}
\end{figure}
A substitution of these solutions into Eqs. \eqref{N_Q-eq1} and \eqref{N_S-eq1} gives a system of equations
\begin{gather}
 \Bigl[-i\omega +  \left(\frac{1}{h\nu_p} + \frac{e^2}{C}\right) (1-e^{-\p_L})\Bigr]  N_Q 
 = 0, 
 %\frac{eV}{2h}\,\left(1 + 2\,\frac{e^2\nu_p}{C}\right) (1-e^{-\p_L})\left[ e^{i\omega x_0/v} - e^{i\omega (L-x_0)/v} \right],
\label{N_Q-eq2}\\
  \Bigl[-i\omega + \frac{1}{\tau_s} + \frac{1-e^{-\p_L}}{h\nu_p}\Bigr]  N_S 
   =\frac{eV}{h}\,(1-e^{-\p_L}). 
   %\left[ e^{i\omega x_0/v} + e^{i\omega (L-x_0)/v} \right],
\label{N_S-eq2}
\end{gather}
which suggests that $N_Q=0$. By substituting the value of $N_S$ from Eq. \eqref{N_S-eq2} into Eqs. \eqref{n--eq2}
and \eqref{I-eq1}, one obtains the expression for the current at the left and right terminals. Using the dimensionless 
spin-flip time
\be
 \eta= \tau_s\,\frac{1 - e^{-\p_L}}{h\nu_p},
 \label{eta}
\ee
it may be presented in the form
\be 
 I_{\omega} = \frac{e^2V}{2h} 
 \left[ 1 + e^{-\p_L} + \frac{ (1 - e^{-\p_L})\,\eta }{ 1 + \eta - i\omega\tau_s} \right].
 \label{I-eq3}
\ee
The real and imaginary parts of the frequency-dependent conductance are shown in Fig. 2.  In the low-frequency limit 
$\omega \ll \tau_s^{-1}$, Eq. \eqref{I-eq3} gives 
\be
 I_{dc} = \frac{e^2V}{2h}\,\frac{1 + 2\eta + e^{-\p_L}}{1 + \eta},
 \label{I_dc}
\ee
which suggests that the dc conductance varies from $e^2/h$ to $e^2/2h$ and increase either with 
decreasing tunnel coupling $\p_L$ or increasing spin-flip rate $\tau_s^{-1}$. The contour plot of this
quantity is shown in Fig. 3 as a function of $\p_L$ and the dimensionless spin-flip rate $h\nu_p/\tau_s$.
In the high-frequency limit, 
it follows from Eq. \eqref{I-eq3} that
\be 
 I_{hf} = \frac{e^2V}{2h}\,(1 + e^{-\p_L}).
 \label{I_hf}
\ee
The high-frequency conductance also varies from $e^2/h$ to $e^2/2h$, but is independent of the spin-flip 
rate and is always smaller than the dc conductance. The high-frequency current is in phase with the ac voltage, 
and the phase shift between them appears only at $\omega\sim\tau_s^{-1}$.

\begin{figure}
  \centering
  \includegraphics[width=0.9\columnwidth]{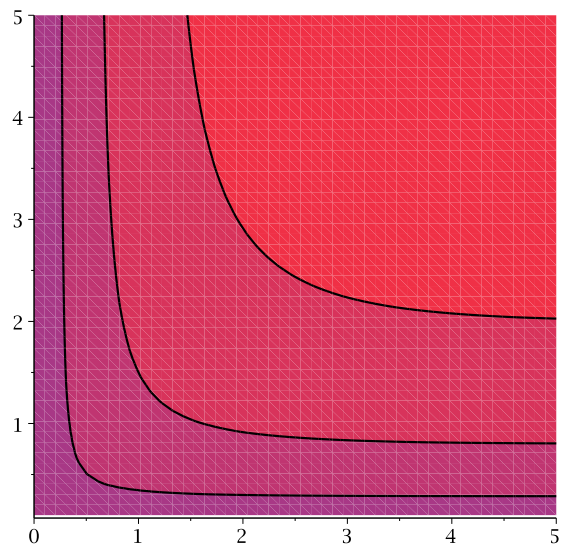}
  \caption{(color online). Contour plot of the dc conductance Eq.~\eqref{I_dc} as a function of $\p_L$ 
  and $h\nu_p/\tau_s$. Brighter colors correspond to smaller values of conductance}
  \label{fig3}
\end{figure}

\section{Discussion}
Though the response is calculated at frequencies much lower than the inverse time of flight of an electron between the
terminals and the pileup of the charge is forbidden in the system, it still exhibits a dispersion related with
spin imbalance in the puddle. At low frequencies, the conductance monotonically decreases as the coupling to the puddle 
and the spin-flip 
rate in it increase. Eventually it becomes equal to one half of the conductance in the absence of the puddle.  This means that 
the puddle breaks the system into  two independent quantum resistors, each with a conductance $e^2/h$. When connected
in series, these resistors exhibit the conductance two times lower, i.~e. $e^2/2h$. Should there be $m$ puddles strongly coupled 
to the edge states, the dc conductance would be $m+1$ times smaller than $e^2/h$. In some sense, increasing the frequency is 
equivalent to increasing the spin-flip rate, and it leads to a similar decrease of conductance. The single-puddle model involves
three unknown parameters, i.~e. $\p_L$, $\nu_p$, and $\tau_s$. All of them can be determined by comparing the experimental
dispersion curve with Eq. \eqref{I-eq3}. If it is not possible to measure the ac response in the whole frequency range, it may be 
possible to measure it in the dc regime and at a frequency well above $\tau_s^{-1}$, so one still can extract $\p_L$ and the product 
$h\nu_p/\tau_s$ by means of Eqs. \eqref{I_dc} and \eqref{I_hf}.

The estimates \cite{Raichev12} show that the Fermi velocity in the edge states of HgTe quantum wells is about 
$5\times10^5$ m/s. If the length of the edge state is one micron, the condition $\omega < v/L$ will be fulfilled up to the
terahertz frequencies. It is more difficult to give reliable estimates of the spin-flip rate in the puddle. In low-temperature
experiments on Au and Cu, the spin-flip time was $\approx 0.1$ ns \cite{Pierre03}. To the best of our 
knowledge, so far the ac response in 2D topological insulators was measured at a constant frequency of 2.5 THz and for 
several-micron long samples \cite{Kvon16}, which is marginal for testing the obtained results. One could extend the frequency 
limits for observing the predicted  
effects by choosing a shorter distance between the measuring probes and making an artificial puddle between them by 
approaching a charged
STM tip or by selective doping. This would provide a test for the proposed model of the conductance suppression in the edge 
states of 2D topological insulators.

\section{Conclusion}
We have calculated a current response to an ac voltage of a pair of edge states in a 2D topological insulators coupled
by tunneling to a conducting puddle in its bulk, where the electrons can flip their spin. Our goal was to provide a means
of experimental detection of such puddles. In a presence of such a puddle, the response exhibits a dispersion at the
inverse spin-flip time in the puddle. Its real part decreases from the zero-frequency value to a smaller value, while
its imaginary part exhibits a maximum at this frequency. By comparing the low-frequency and high-frequency response,
one can determine the parameters of the puddle.

\begin{acknowledgements}
This work was supported by Russian Science Foundation under Grant No. 16-12-10335.
\end{acknowledgements}

\end{document}